# Incoherent Light-Driven Nonlinear Optical Extreme Learner via Data Reverberation

Bofeng Liu, Xu Mei, Sadman Shafi, Tunan Xia, Iam-Choon Khoo, Zhiwen Liu, and Xingjie Ni*

*Department of Electrical Engineering, The Pennsylvania State University, University Park, PA 16802, United States*

* xingjie@psu.edu

Artificial neural networks have revolutionized fields from computer vision to natural language processing, yet their growing energy and computational demands threaten future progress. Optical neural networks promise greater speed, bandwidth, and energy efficiency, but suffer from weak optical nonlinearities. Here, we demonstrate a **low-power, incoherent-light-driven** optical extreme learner that leverages "data nonlinearity" from optical pattern reverberation, eliminating reliance on intrinsic nonlinear materials. By encoding input data in the spatial polarization distribution of a tailored optical cavity and allowing light to pass through it multiple times, we achieve nonlinear transformations at extremely low optical power. Coupled with a simple trainable readout, our optical learner **consistently outperforms linear digital networks** in standard image classification tasks and XOR benchmarks, delivering accuracy **matching fully nonlinear digital models**. Our compact, energy-efficient approach significantly reduces complexity, cost, and energy consumption, paving the way for practical, scalable all-optical machine learning platforms.

## Introduction

Optical computing has its roots in the 19th century, with Ernst Abbe's pioneering work on Fourier optics (*1*) and has captivated scientific imagination since Duffieux systematically introduced Fourier integrals into optics in the 1940s (*2-6*). Despite early enthusiasm, the rapid advancement of digital electronics and the subsequent explosion of general-purpose computing overshadowed optical computing(*7, 8*). However, in recent years, optical computing has seen a notable resurgence, driven by its inherent advantages in parallel processing capability, low latency, and significantly reduced power consumption (*9-17*). This renewed interest is particularly evident in applications targeting artificial neural network (ANN) acceleration (*10, 12, 18-31*). Optical neural network (ONN) architectures uniquely offer potential solutions to fundamental challenges faced by electronic neural networks, such as speed bottlenecks and escalating power consumption. These advantages are becoming increasingly critical as global artificial intelligence (AI) workloads continue their exponential growth trajectory (*32-34*). Optical methods, therefore, represent a promising frontier for overcoming the limitations inherent in traditional electronic computing paradigms.

A fundamental requirement for artificial neural networks (ANNs) to serve effectively as universal approximators is the integration of nonlinear activation functions (*35-38*). In ONNs, while optical nonlinearities can be introduced through various means, including leveraging intrinsic optical material nonlinearities such as Kerr effect (*39-42*), exploiting nonlinear responses in detector (*43, 44*), or employing other specialized nonlinear optical media such as saturable absorbers (*18)*, and laser-cooled atoms with electromagnetically induced transparency (*19, 45*). Nevertheless, each of these approaches has notable limitations, including the necessity of high-power laser sources, slow response times, or complex and costly fabrication processes (*46*).

Recently, an alternative approach has been proposed wherein input data undergoes repeated scattering interactions within data bearing structures, inducing nonlinear relationships between the scattered optical fields and the input data (*17*). This concept has been demonstrated through multiple scatterings involving a digital mirror device (DMD) coupled with an integration sphere (*47*) , as well as through interactions between a spatial light modulator (SLM) and a mirror (*48*). These systems generate speckle patterns whose nonlinear characteristics correlate with the input data and are processed in conjunction with a simple digital neural network to achieve tasks such as image classification. However, those systems experimentally only achieve relatively low classification accuracy, underperforming even basic linear digital networks. Additionally, these methods rely on coherent input illumination, inevitably involving laser sources that tend to incur higher energy consumption and overall system costs. The need to utilize components like DMDs and SLMs are inherently complex and expensive, limiting the practicality and scalability of these systems for widespread applications.

Here, we introduce **an optical extreme learner specifically designed to operate using only low-power incoherent light, consistently surpassing linear digital neural** **networks, reaching performance levels comparable to established nonlinear neural networks in diverse tasks such as image classification and nonlinear image processing.** An extreme learner, a.k.a., an extreme learning machine, is a neural network featuring nonlinear hidden nodes with randomly assigned, fixed parameters and a single trainable output layer. Our optical extreme learner employs a small pixelated transparent liquid crystal display (LCD) panel that modulates transmitted light amplitude, sandwiched between two partially reflective interfaces, with a compact total device thickness of only 1.88 mm. Input data are encoded onto the LCD panel and illuminated by a uniform-intensity incoherent light beam. The repeated transmission of light through the LCD panel generates a nonlinear mapping from input data to output signals. These outputs are digitally captured and processed through a single-layer, trainable linear readout network. Employing our optical extreme learner, **we attained image classification accuracies 96.82%, 98.20%, and 81.21% – results that is unattainable by purely linear neural networks** – on standard datasets (MNIST, CMNIST, and EMNIST) **with white light**. Additionally, we **successfully demonstrated XOR operations** on input images – a task inherently challenging for linear networks. Our approach provides a promising foundation for developing low-power, compact, and cost-effective optical processors without the need of optical nonlinearity or laser sources. We believe this technology holds significant potential to accelerate current AI applications, offering superior computational speed and improved energy efficiency.

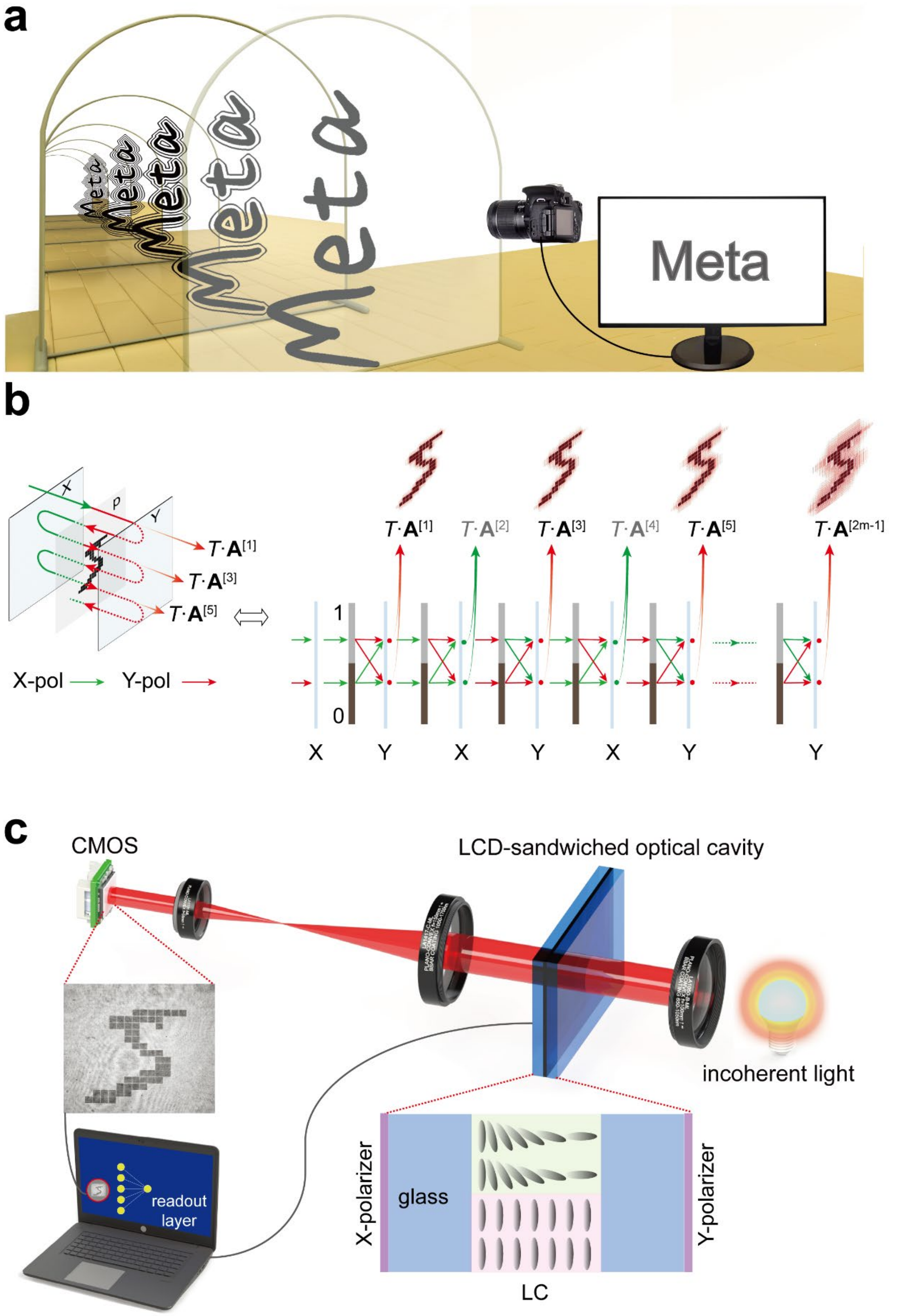

**Figure 1. Schematic diagram of the optical extreme learner with multi-pass data reverberation.** (**a**) Conceptual illustration showing how data is repeatedly multiplied by itself within an "infinity mirror" configuration. (**b**) Practical implementation of the optical cavity

formed by two orthogonally oriented polarizing mirrors (X and Y) and a transparent (90°-twist planar-aligned, pixelated) liquid crystal display (LCD) panel. The decomposition of each reflection and transmission is shown: The incident beam first passes through an X-oriented polarizer, acquiring a defined polarization state, which is subsequently modulated by the liquid crystal layer with an input pattern **p**. Upon diffraction from the exit LC cell window, the Y-polarized component of the beam is selected by the Y-oriented polarizer, resulting in the first transmitted electric field component, denoted as $T\cdot\mathbf{A}^{[1]}$. Through successive internal reflections and polarization modulations within the system, additional transmitted components such as $T\cdot\mathbf{A}^{[3]}$, $T\cdot\mathbf{A}^{[5]}$, etc., are generated. (**c**) Experimental setup of our optical extreme learner: a collimated beam illuminates the LCD-sandwiched cavity, and a grayscale CMOS camera records the resulting nonlinear patterns by focusing on the Y plane. The optical cavity is comprised of polarizers (X and Y), glass, and liquid crystal (LC).

## Working Principle

Our optical extreme learner consists of a pair of partial mirrors and a pixelated intensity modulation layer – specifically, we use a small transparent LCD panel sandwiched between two polarizing partial mirrors (Fig. 1c). Its working principle can be conceptualized as an input data-bearing structure placed within an infinity mirror configuration (Fig. 1a). When an input structure – a spatial transmittance distribution encoding the input data – is positioned between two partially reflective mirrors, a light beam passing through it acquires an intensity pattern modulated by the encoded information. As the modulated beam propagates, diffraction slightly alters its intensity distribution. Upon reflection by the mirrors, the beam repeatedly traverses the transmittance structure, each traversal further modulating the spatial intensity distribution. This data reverberation process – consisting of repeated diffraction, reflection, and transmission cycles – results in the interactions between the input data and its own diffracted field, ultimately generating a nonlinear mapping between the input data and the output optical field.

To clearly illustrate this iterative process, we can "unfold" the infinity mirror configuration into a linear sequence of multiple identical layers, each modulated by the same input data distribution, **p** (Fig. 1b). Here, ***a pair of orthogonally oriented polarizers act as polarization-sensitive partial mirrors***. When the polarization of impinging light aligns with a polarizer's transmission axis, it is highly transmissive with transmittance $T$; conversely, when the polarization is perpendicular to the polarizer's transmission axis, it is attenuated and partially reflected with reflectance $R$. Initially, the incident light has a uniform intensity distribution. After passing through the first polarizer, the spatial polarization distribution of the beam is modulated by LCD pixels, where the input data is encoded. A bright pixel representing a value "1" rotates the polarization, whereas a dark pixel representing a value "0" allows the beam to pass without polarization rotation. The modulated beam then diffracts toward the second polarizer. As the second polarizer's transmission axis is orthogonal to that of the first polarizer, only the portion of the beam with polarization rotated can pass through, producing an intensity pattern $\mathbf{A}^{[1]}$. The unrotated portion of the beam is partially reflected by the second polarizer, entering the second layer, where it undergoes an identical polarization modulation, resulting in intensity pattern $\mathbf{A}^{[2]}$. After traversing the $n^{\text{th}}$ layer, the final output light produced is $\mathbf{A}^{[n]}$. This iterative process can be mathematically described as,

$$\mathbf{A}^{[n]} = \{\mathbf{P} \otimes \boldsymbol{\sigma}_x(\mathbf{A}^{[n-1]}R \circledast \overleftrightarrow{\mathbf{G}})\boldsymbol{\sigma}_x + [(\mathbf{I}-\mathbf{P}) \otimes \mathbf{I}_2](\mathbf{A}^{[n-1]}R \circledast \overleftrightarrow{\mathbf{G}})\} \circledast \overleftrightarrow{\mathbf{G}} \quad (1)$$

where $\overleftrightarrow{\mathbf{G}}$ denotes the Green's function characterizing free-space propagation between two consecutive layers, $\otimes$ indicates the Kronecker product, and $\circledast$ represents convolution. The matrix $\mathbf{P}$ is obtained by diagonalizing the pattern vector $\mathbf{P} = \text{diag}(\mathbf{p}) \in \mathbb{R}^{N \times N}$, $N$ denotes the dimension of the pattern vector, $\mathbf{I} \in \mathbb{R}^{N \times N}$, $\mathbf{I}_2 \in \mathbb{R}^{2 \times 2}$ are identity matrices, and $\boldsymbol{\sigma}_x$ is the Pauli matrix. Details can be found in Supplementary Note 5. The final output pattern of the system is the sum of the intensity distributions emerging from all odd-numbered interfaces (equivalent to polarizer Y in Fig. 1b), which can be expressed as

$$\mathbf{A}_{\text{out}} = \textstyle\sum_m T \cdot \mathbf{A}^{[2m-1]} \quad (2)$$

where $m$ is a positive integer. From Eq. (2), it is evident that the output pattern comprises a series multiplicative terms involving interactions with the input data, thereby establishing a nonlinear relationship between the input and output. Each passage of light through the LCD pixels multiplies the input data pattern, effectively increasing the nonlinear order of the data by one. The maximum achievable nonlinear order, determined by the effective number of reflections within the optical cavity, can be controlled by adjusting the reflectance coefficients of the mirrors. The resulting nonlinearly mapped output intensity pattern is captured by a standard CMOS camera positioned after the second mirror. The captured data is then processed by a simple, trainable linear network consisting of only a single readout layer, ensuring minimal computational load.

In contrast to conventional material-based nonlinearities, which rely on intrinsic optical properties, the nonlinear relationship between the input and output fields in our system arises from passive multiple interactions among spatially distributed pixels on the LCD. This mechanism is inherently independent of input light power and coherence, enabling nonlinear transformations even with low-power continuous-wave and incoherent light sources. Consequently, our approach offers significantly improved energy efficiency compared to traditional nonlinear optical systems.

## Results

Each pixel of the LCD panel used in our optical extreme learner measures 370×420 $\mu m^2$, with a gap of 20 μm between adjacent pixels. A micro-controller is employed to drive the LCD panel. Our experiments utilized both coherent and incoherent light sources: a coherent 633-nm Helium-Neon laser and a temporally incoherent white supercontinuum source filtered by a 40-nm bandpass centered at 650 nm. The collimated input beam passed through the optical cavity, and the resulting output pattern was captured by a monochrome camera positioned downstream. Details of the experimental setup are described in the Materials and Methods section.

To systematically evaluate our optical extreme learner, we employed the widely used

MNIST dataset, consisting of 60,000 training images and 10,000 testing images. Due to field of view limitations of our optical setup, we down-sampled the original MNIST images from 28×28 pixels to 20×20 pixels. The experimentally obtained nonlinearly transformed outputs were then processed through a single-layer linear readout network to get the classification results.

We also simulated the optical processes within our compact cavity using the beam propagation method (BPM). During simulations, MNIST images were pixelated to closely resemble their appearance on the LCD panel. The resulting simulated output patterns exhibited additional fine structures beyond the original images – an effect also observed in experimentally captured images (Fig. 2c). These features are attributed to the combined effects of diffraction and data reverberation, creating nonlinear transformations of the input data.

For comparison, two baseline scenarios using linear digital neural networks were established. The first baseline involved directly inputting original MNIST images into a linear neural network to establish the linear performance limit. The second baseline employed pixelated MNIST images to confirm that performance improvements were not merely due to image pixelation.

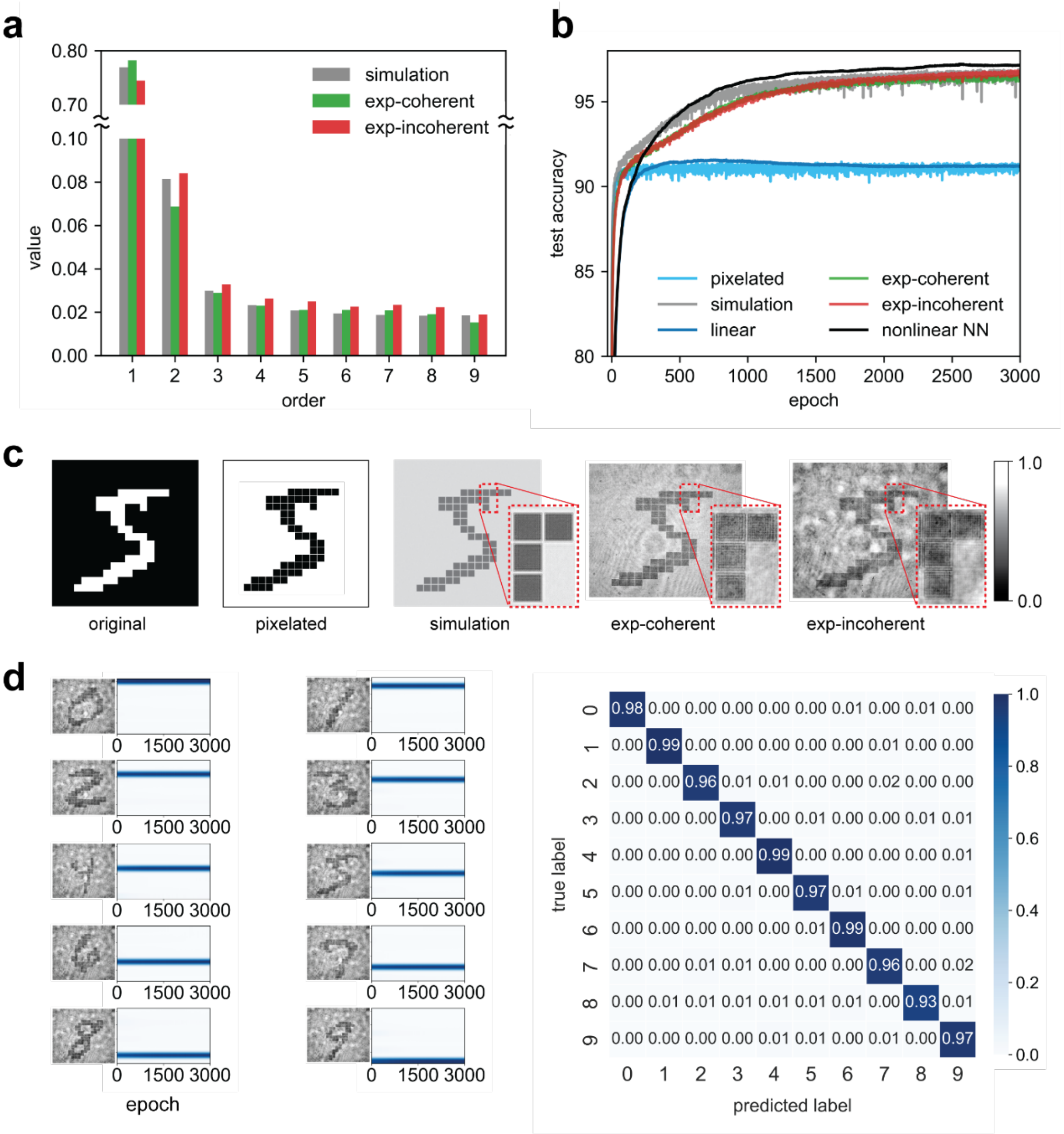

**Figure 2. Performance analysis of the optical extreme learner on MNIST dataset.** (**a**) Distribution of nonlinear orders derived via Boolean analysis. “exp-coherent” and “exp-incoherent” indicate experiments using laser and white-light sources, respectively, showing minimal differences in higher-order nonlinear coefficients. (**b**) Blind test accuracy curves over training epochs. The optical extreme learner driven by a white-light source (red line) achieves 96.82% accuracy, while driven by a laser source (green line) achieves 96.54%. For comparison, a linear neural network (blue line) reaches 91.50%, the simulated optical learner (gray line) attains 96.88%, and a ReLU-based nonlinear neural network (black line) achieves 97.21%. (**c**) Example images used in the tests (from left to right): Original MNIST image, pixelated MNIST

image, simulated MNIST output image, experimentally obtained image using coherent illumination, and experimentally obtained image using incoherent illumination. The optical process encodes complex interactions between the diffracted input pattern and the original encoded data through multi-passes of the beam through the LCD, thereby capturing subtle inter-pixel interactions within the resulting intensity patterns. (**d**) Left panel: convergence curves for different class images. Right panel: the confusion matrix of blind test.

Experimental results from our optical extreme learner demonstrated classification accuracies of 96.54% (coherent illumination) and 96.82% (incoherent illumination), closely matching the simulation accuracy (within approximately 0.06%). These results notably exceed the performance limit of purely linear networks (~91.50% in our tests, similar to that reported in (*49*)) by more than 5%, approaching the performance of fully nonlinear digital neural networks (97.21%). Additionally, training the linear readout layer consistently demonstrated rapid and robust convergence across different image classes (Fig. 2d). These findings clearly indicate that nonlinear interactions arising from input data reverberation in our optical system enable classification performance beyond the capabilities of linear neural networks.

Furthermore, to quantitatively assess the nonlinearity introduced by our optical system, we applied Boolean analysis (*17, 50*) to extract the expansion coefficients corresponding to different nonlinear orders of the input data present in the output patterns. Using binary inputs of 3×3 pixels (Supplementary Note 1), we evaluated the distribution of these coefficient, $\mathbf{C}_k$, in both simulated and experimentally measured outputs. Results from both consistently exhibit the presence of higher-order nonlinear terms (Fig.2a). These findings confirm that our optical structure generates nonlinearity for the input data effectively, which plays a vital role in enhancing system performance.

To further validate the generalizability of our optical extreme learner and its superiority over linear networks, we tested its performance on additional datasets: Chinese MNIST (CMNIST) dataset, comprising 15,000 images of handwritten Chinese characters across 15 classes, and the Extended MNIST Letters (EMNIST) dataset, which includes 145,600 images of uppercase and lowercase letters across 26 classes. It is important to note that, unlike the MNIST and EMNIST datasets, the CMNIST dataset does not provide a predefined division into training and testing subsets, complicating standard evaluation procedures. To address this issue, we employ a 10-fold cross-validation approach – a widely-used method to minimize data selection bias (*51, 52*). In this method, the entire dataset is randomly partitioned into ten equally sized folds. During each iteration, one fold is held out for validation, while the remaining nine folds are used for training (Supplementary Note 3). This procedure guarantees that every sample serves exactly once as a validation data point and participates in training across all other iterations. By averaging the performance metrics across all folds, we achieve a robust and unbiased estimation of the model's accuracy, effectively mitigating potential biases introduced by arbitrary data splits.

Experimental results (Fig. 3) show a significant improvement in classification accuracy due to the integration of optical nonlinear mapping. This improvement is evident in both the 15-class CMNIST task and the 26-class EMNIST task. Specifically, for CMNIST, classification accuracy increased from 43.64% (linear network) to 98.20% with incoherent light and 98.19% with coherent light – both even exceeding the performance of a fully nonlinear digital neural network (68.82%). For EMNIST, accuracy improved from 70.84% (linear network) to 81.21% (incoherent) and 85.18% (coherent). These advantages are further supported by confusion matrices, which show most predictions concentrated along the diagonal, indicating high classification accuracy (Fig. 3b). One exception is a notable confusion between the letters ‘i’ and ‘l’ in EMNIST. This misclassification can be attributed to their inherent visual similarity, compounded by down-sampling effects that sometimes eliminate the distinguishing dot above the letter ‘i.’ Despite this minor challenge, our results demonstrate that the optical extreme learner delivers robust and superior classification performance in blind prediction tests across different datasets, significantly outperforming purely linear network architectures.

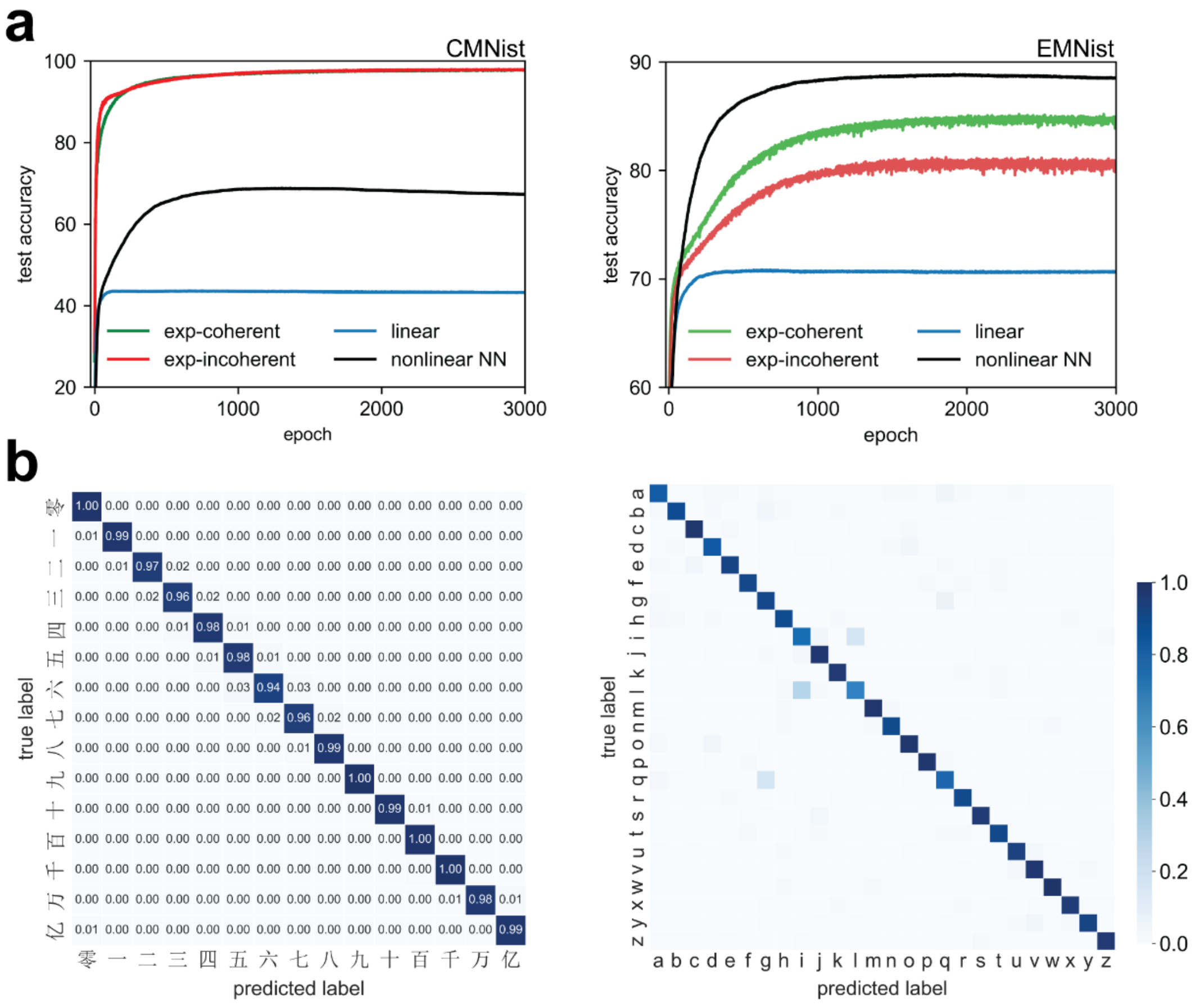

**Figure 3. Performance analysis of the optical extreme learner on CMNIST and EMNIST datasets.** (**a**) Blind test accuracy curves over training epochs. Left panel (CMNIST): Comparison of optical extreme learner performance under white-light illumination (red line, accuracy: 98.20%) and laser illumination (green line, accuracy: 98.19%), against a purely linear neural network (blue line, accuracy: 43.64%) and a nonlinear neural network with ReLU activation (black line, accuracy: 68.82%). Right panel (EMNIST): Comparison of optical extreme learner performance under white-light illumination (red line, accuracy: 81.21%) and laser illumination (green line, accuracy: 85.18%), against a purely linear neural network (blue line, accuracy: 70.84%) and a nonlinear neural network with ReLU activation (black line, accuracy: 88.86%). (**b**) Confusion matrices of blind tests for CMNIST (left panel) and EMNIST (right panel) datasets, highlighting the excellent classification performance of the optical extreme learner.

Our optical extreme learner is not limited to image classification tasks, it is also applicable to other computational problems. A historically significant challenge in neural networks is the XOR problem, which early single-layer perceptrons were unable to solve. This limitation contributed to the AI winter of the 1960s (*53, 54*). The introduction of multilayer neural networks with nonlinear activation functions later overcame this barrier (*55*), sparking renewed interest and progress in the field – often referred to as the second spring of neural networks. To demonstrate the computational capability of our optical extreme learner in this context, we applied it to perform XOR operations on the EMNIST dataset. Specifically, we selected two regions within each image with a slight spatial offset, as highlighted by the red and green boxes in the top row of Fig. 4a and applied an XOR operation between these two regions. We conducted this task using four different configurations: a digital nonlinear neural network with ReLU activation, a purely linear network, and our optical extreme learner under both coherent and incoherent illumination.

Figure 4a presents sample XOR outputs from the training and testing datasets across all four configurations. The results clearly show that both the digital nonlinear network and the optical extreme learner (under both illumination conditions) produce outputs that closely match the ideal XOR operation. In contrast, the linear network fails to generate meaningful XOR patterns. For quantitative comparison, we calculated the Structural Similarity Index (SSIM) between each output and the digitally computed ground truth (Fig. 4b). The digital nonlinear network and both optical extreme learner configurations achieved SSIM values close to 1.0, indicating high structural similarity. In contrast, the linear network achieved an SSIM of only 0.45, reflecting poor XOR performance.

These results provide compelling evidence that the optical extreme learner effectively addresses the classical XOR problem – long considered a benchmark for demonstrating nonlinearity in neural systems. This success further validates the system's capability to perform nonlinear operations and highlights the strength of optically induced data nonlinearity in enhancing computational power.

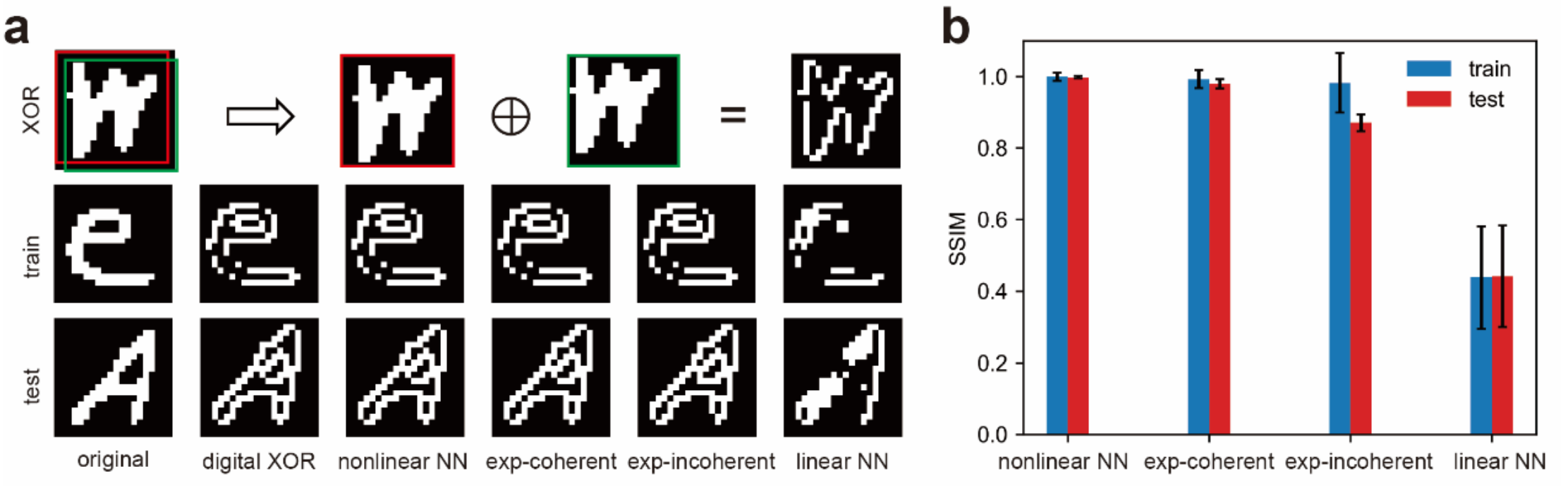

**Figure 4. Performance analysis of the optical extreme learner on the image XOR operation.** (**a**) The first row illustrates the XOR computational process. It shows two selected regions (highlighted by red and green boxes) from the original images, and their corresponding XOR result, forming a hollow character. The second and third rows display training and testing character sets evaluated by different neural networks. The first two columns present the original images and corresponding XOR ground truth. Columns labeled "nonlinear NN," "exp-coherent," "exp-incoherent," and "linear NN" represent XOR results obtained from a nonlinear neural network, our optical extreme learner under coherent (laser) and incoherent (white-light)

illumination, and a linear neural network, respectively. (b) Average Structural Similarity Index Measure (SSIM) comparing digitally computed XOR results to those obtained via: a nonlinear neural network, our optical extreme learner experiments under coherent and incoherent illumination, and a linear neural network.

In the first row, select two different parts of the picture in the red and green boxes to do XOR computing. The XOR result is a hollow character. The second and third rows are the train and test characters by different neural networks, where original pictures and the XOR ground truth are in the first two columns. The 'nonlinear NN' represents the results from a nonlinear neural network. The 'exp-coherent' and 'exp-incoherent' are XOR computing from experimental data under coherent and incoherent light sources. The 'linear NN' and 'nonlinear NN' indicates the XOR results from a linear and nonlinear neural network, respectively. (b) The average SSIM between the digitally computed XOR results with those obtained using a nonlinear neural network, experimentally obtained using our optical extreme learner under coherent and incoherent light sources, and a linear neural network, respectively.

**Table 1. Test accuracies and SSIM values for all evaluated methods.**

| **Method** | **Classification accuracy MNIST (%)** | **Classification accuracy CMNIST (%)** | **Classification accuracy EMNIST (%)** | **XOR Training SSIM** | **XOR Testing SSIM** |
|---|---|---|---|---|---|
| **Linear digital neural networks** | 91.50 | 43.64 | 70.84 | 0.438 | 0.446 |
| **Fully nonlinear digital neural networks** | 97.21 | 68.82 | 88.86 | 0.999 | 0.998 |
| **Our optical learner w/ *coherent* light** | 96.54 | 98.19 | 85.18 | 0.993 | 0.980 |
| **Our optical learner w/ *incoherent* light** | 96.82 | 98.20 | 81.21 | 0.982 | 0.870 |

*Discussion*

We successfully demonstrated an optical extreme learner - consisting of a compact optical cavity with two partial reflective mirrors and a liquid crystal screen only having a thickness of 1.88 mm. Our device exploits the nonlinear interplay between the input data and the output optical field, created by multiple passes of light reflection and diffraction – each pass multiplying the spatial transmittance distribution encoded by the input. Coupled with a simple, single-layer linear readout, our system delivers strong performance on a variety of tasks, including image classification with multiple commonly used benchmark datasets (MNIST, CMNIST, and EMNIST), as well as image XOR operations – a known challenge for purely linear networks. Both numerical simulations and experimental results confirm that the optical extreme learner surpasses linear networks across all tested tasks, matching the performance of fully nonlinear digital architectures (see **Table. 1**)

A key advantage of our approach lies in its reliance on intensity-based modulation rather than phase modulation, making it inherently compatible with incoherent light sources. **Our experiments show that using a low-power incoherent source yields results nearly identical to those achieved with a coherent laser, removing the need for more costly or high-power laser illumination.** In addition, the system's degree of nonlinearity can be adjusted by tweaking mirror reflectivity. Furthermore, the input pattern is provided by a low-cost LCD, which makes our system much more cost-effective compared to other optical computing implementations where expensive DMD or SLM have usually been used. (*26, 40, 56, 57*). Looking ahead, our optical extreme learner can be readily integrated with optical diffractive neural networks (ODNN) (*56*), such as those based on metasurfaces, to create deeper, more sophisticated all-optical computing architectures.

Additionally, in contrast to traditional approaches that typically depend on lower-order material nonlinearities or detector nonlinearities, our method harnesses higher-order nonlinearities through data reverberation. Boolean analysis of the optical output reveals the presence of numerous nonlinear terms beyond third order, significantly expanding the dimensionality of the input data space. **The utilization of these higher-order nonlinearities improves data separability by enhancing computational expressiveness (*58*) and facilitating more comprehensive feature-space expansion (*59*), thereby enabling superior generalization and adaptability in complex computational tasks.**

In summary, our optical extreme learner, consisting of a 1.88-mm-thick cavity combined with a simple linear readout, operates seamlessly under incoherent white-light illumination. It exceeds the performance limit of purely linear neural networks and attains accuracy comparable to fully nonlinear digital systems marking a vital step forward for fast, low-latency, energy-efficient optical computing solutions that preserve the full capabilities of nonlinear neural networks.

**Acknowledgments**

The authors acknowledge valuable discussions with T.-H. Lin.

**Funding:** The work was partially supported by the Air Force Office of Scientific Research under award number FA2386-24-1-4054 and the National Science Foundation under grant agreements ECCS-2047446.

**Author contributions:** Conceptualization: XN, ICK, and ZL. Methodology: BL, XM, SS, and XN. Investigation: BL, XM, and XN. Visualization: BL, XM, TX, and XN. Supervision: XN. Writing – original draft: BL. Writing review & editing: BL, XN, ICK, and ZL. All authors read and approved the manuscript.

**Competing interests:** Authors declare that they have no competing interests.

**Data and materials availability:** All data are available in the main text or the supplementary materials.

**Supplementary Materials**

Materials and Methods

Supplementary Text

Figs. S1 to S5

References (S1–S9)

## Materials and Methods

### *Data preparation and neural network training*

To evaluate the optical extreme learner's performance, we used the MNIST (*60*), EMNIST (*61*), and Chinese MNIST (*62*) datasets. The MNIST dataset comprises 70,000 images of handwritten digits (0–9), each with dimensions of 28×28 pixels. MNIST contains 70,000 handwritten-digit images (0–9) at 28 × 28 px. EMNIST provides 145,600 handwritten-letter images spanning 26 classes ('A/a'–'Z/z'), also at 28 × 28 px. Chinese MNIST offers 15,000 handwritten-numeral images (values 0–108) originally at 64 × 64 px. All images were down-sampled to 20 × 20 px with bilinear interpolation to match our optical system's field of view. Every classification and XOR results reported here use these down-sampled datasets.

In our experiments, MNIST dataset was partitioned into a training set of 60,000 images and a test set of 10,000 images; EMNIST dataset was divided into 124,800 training images and 20,800 test images; and the Chinese MNIST dataset was split into 13,500 training images and 1,500 test images in each fold of cross-validation. All neural networks, comprising a single linear readout layer, were implemented using the PyTorch framework. Cross-entropy loss function was used for classification, whereas mean-squared error (MSE) loss function was applied in XOR experiments.

We employed the Adam optimizer for network training, setting the learning rate to $1\times10^{-5}$ to ensure stability. The batch sizes were 1200, 2400, and 500 for MNIST,

EMNIST, and Chinese MNIST datasets, respectively. We reshuffled the training data at every epoch to reduce bias and improve generalization. All runs were executed on a Linux workstation equipped with an NVIDIA GeForce RTX 4090 (24 GB).

***Experiment setup***

A collimated beam illuminates a compact binary LCD panel that served as the programmable input in our experiment (Fig. 1c). Illumination begins with light delivered through a 10 × objective (PlanC N 10×, Olympus) into a single-mode fiber (P1-630A-FC-5, Thorlabs). The fiber output is collimated by a fiber collimator and a 150 mm lens. The 2.2″ transparent LCD panel (Crystalfontz), with a pixel pitch of 420 μm × 370 μm, was positioned orthogonally to the incident beam and driven by an Arduino Nano microcontroller via serial communication. Nonlinear modulation arises inside the compact LCD cavity, which comprises two glass plates enclosing the liquid-crystal layer and carries crossed polarizers on the glass plates' outer surfaces. Multiple reflections, diffraction, and self-interaction of the intensity-modulation pattern within the liquid crystal combine to generate the desired high-order nonlinear response. The resulting optical field at the LCD plane was imaged onto a monochrome CMOS camera (DMK 33GX265, Imaging Source) through a 150-mm lens and a 50-mm lens configured as a telescope setup. To compare coherent and incoherent operation, we alternated between a 633 nm He-Ne laser and a supercontinuum source (SuperK COMPACT, NKT Photonics) passed through a 40 nm band-pass filter centered at 650 nm (FBH650-40, Thorlabs).